\documentclass[11pt]{article}
\usepackage{tikz}
\usetikzlibrary{arrows.meta,positioning}
\usepackage{graphicx} % preamble に
\usepackage{amsmath,amssymb}
\usepackage{geometry}
\usepackage{hyperref}
\usepackage{graphicx}

\newcommand{\GL}{\mathrm{GL}}
\title{Trilinear Kernel Structure and Its Gravitational Realization}
\author{
Takeshi Fukuyama\\[2mm]
{\small Research Center for Nuclear Physics (RCNP), Osaka University, Ibaraki, Osaka 567--0047, Japan}
}
\begin{document}
\maketitle

\begin{abstract}
We clarify the structural role of trilinear kernels in multidimensional
integrable hierarchies and in stationary axisymmetric gravity.
The Yu--Toda--Fukuyama (YTF) trilinear equation of
Ref.~\cite{YuTodaSasaFukuyama:1998hierarchy} is shown to represent not a
particular evolution equation but a universal kernel that generates the
entire $(3+1)$--dimensional hierarchy by selecting commuting flows.
The frequently quoted trilinear equation of
Ref.~\cite{YTSF1998} is identified as one such flow of this kernel.

We further show that stationary axisymmetric gravity corresponds to a
projective realization of the YTF kernel rather than to any single flow.
Imposing $\GL(2)$ covariance and homogeneity on the kernel leads uniquely
to a gravitational trilinear kernel $\mathcal{Y}(\tau_0,\tau_1)$, whose
vanishing reproduces the Ernst equation.
The Tomimatsu--Sato family \cite{Tomimatsu1972} and related bilinear solutions are shown to arise
as degenerate submanifolds of this projected trilinear structure, in
agreement with the multilinear analysis of
Ref.~\cite{Fukuyama:2025TS}.

These results establish a unified structural framework linking
multidimensional trilinear integrability, stationary gravity, and bilinear
solution sectors, and clarify why trilinear kernels are both necessary and
sufficient for describing soliton dynamics with projective geometry.
\end{abstract}

%%%%%%%%%%%%%%%%%%%%%%%%%%%%%%%%%%%%%%%%%%%%%%%%%%%%%%%%%%%%
%%%%%%%%%%%%%%%%%%%%%%%%%%%%%%%%%%%%%%%%%%%%%%%%%%%%%%%%%%%%
\section{Introduction}
This paper addresses a basic structural problem in integrable systems:
Hirota's bilinear formalism provides a complete and remarkably effective
description in $(1+1)$ dimensions, but it ceases to be structurally
sufficient in higher dimensions, where genuinely multidirectional
interactions appear and a minimal extension is required.
Our viewpoint is that the obstruction is not merely technical but
conceptual: bilinear identities encode only \emph{pairwise} interference
between two copies of a single $\tau$--function.
In genuinely multidimensional settings, however, solitons can interfere
simultaneously in three independent directions, and such irreducible
three--way interference requires a framework beyond bilinear theory.

A remarkable aspect of the present problem is that this minimal extension
is realized in a concrete physical system.
Stationary axisymmetric gravity admits a projective $\tau$--representation,
and the associated $\GL(2)$ covariance makes it intrinsically a
two--component system.
We will show that, once one insists on homogeneity and projective
covariance, the natural integrable object is not a bilinear identity but a
trilinear \emph{kernel}, and that the universal Yu--Toda--Fukuyama kernel
provides precisely the organizing principle for multidimensional flows.
Figure~\ref{fig:scheme-eq55} summarizes the resulting hierarchy:
\[
\text{(bilinear sector)}\ \Rightarrow\ \text{(trilinear kernel/flows)}
\ \Rightarrow\ \text{(gravitational projection)}\ \Rightarrow\
\text{(bilinear degenerations)} .
\]
Here the last arrow does not mean a return to ordinary $(1+1)$--dimensional
integrability.
It indicates that, within the projective gravitational realization of the
trilinear kernel, there exist special degenerate submanifolds on which the
three--way interference collapses and the dynamics becomes effectively
bilinear again.
This is precisely what happens in the Tomimatsu--Sato family and related
bilinear solution sectors discussed in Sec.~\ref{sec:TS}.

In the first half of the paper we discuss this chain as a problem of
integrable structure.
In the second half we show that it is realized in a concrete physical
system: stationary axisymmetric gravity.
The crucial point is that gravity is naturally described by a \emph{projective}
$\tau$--representation, so that a single--$\tau$ bilinear identity is not
the appropriate starting point.
This is precisely the setting in which a trilinear kernel becomes the
minimal homogeneous object.

%%%%%%%%%%%%%%%%%%%%%%%%%%%%%%%%%%%%%%%%%%%%%%%%%%%%%%%%%%%%
\paragraph{Stationary gravity as a projective system.}
Stationary axisymmetric vacuum gravity is governed by the Ernst equation
\cite{Ernst1968}
\begin{equation}
(\Re\,\mathcal{E})
\left(\partial_\rho^2+\frac{1}{\rho}\partial_\rho+\partial_z^2\right)\mathcal{E}
=(\partial_\rho\mathcal{E})^2+(\partial_z\mathcal{E})^2 ,
\label{Ernst}
\end{equation}
whose nonlinearity reflects the genuine self--interaction of the
gravitational field.
The Ernst potential is represented projectively by two $\tau$--functions,
\begin{equation}
\mathcal{E}=\frac{\tau_1}{\tau_0},
\qquad (\tau_0,\tau_1)\sim g(\tau_0,\tau_1),\quad g\in \GL(2),
\label{projective}
\end{equation}
so that gravity is intrinsically a two--component, projective system.
This projective nature is the reason that ``a single $\tau$'' is not the
fundamental object for gravity: the physical content resides in the
$\GL(2)$--equivalence class of pairs $(\tau_0,\tau_1)$.
As we will see, the minimal multilinear structure compatible with this
projective geometry is naturally trilinear.

%%%%%%%%%%%%%%%%%%%%%%%%%%%%%%%%%%%%%%%%%%%%%%%%%%%%%%%%%%%%
\paragraph{Bilinear completeness in $(1+1)$ dimensions and its limitation.}
In contrast, $(1+1)$--dimensional integrable equations are successfully
described by Hirota's bilinear formalism.
The basic building block is the bilinear Hirota derivative \cite{Hirota2004},
\begin{equation}
D_x^m D_t^n\, f\cdot g
:=\left.(\partial_x-\partial_{x'})^m(\partial_t-\partial_{t'})^n
f(x,t)\,g(x',t')\right|_{x'=x,\ t'=t}.
\label{eq:HirotaD}
\end{equation}
A bilinear equation has the schematic form
\begin{equation}
\mathcal{B}(D_x,D_t,\ldots)\,\tau\cdot\tau=0.
\label{bilinear}
\end{equation}
Here $\mathcal{B}(D_x,D_t,\ldots)$ denotes a polynomial in Hirota
bilinear derivatives $D_x,D_t,\ldots$ whose vanishing encodes a specific
integrable equation.

A standard example is the KdV equation,
\begin{equation}
(D_x^4+D_xD_t)\,\tau\cdot\tau=0,
\label{KdV}
\end{equation}
with $u=2\partial_x^2\ln\tau$.
Here all nonlinear interactions are encoded through \emph{pairwise}
interference between two copies of the same $\tau$--function.

This pairwise structure explains why bilinear equations are sufficient in
$(1+1)$ dimensions: multi--soliton scattering factorizes into two--body
processes.
However, it also indicates a fundamental limitation.
In higher dimensions, solitons can interfere simultaneously in three or
more independent directions, and such genuine multidirectional
interference cannot be reduced to a sum of pairwise interactions.
This is precisely the point where a trilinear extension becomes
structurally inevitable.

%%%%%%%%%%%%%%%%%%%%%%%%%%%%%%%%%%%%%%%%%%%%%%%%%%%%%%%%%%%%
\paragraph{What this paper does.}
Motivated by this structural obstruction, we organize the multidimensional
extensions of bilinear hierarchies by the trilinear kernel discovered by
Yu, Toda and Fukuyama and by the associated generation of commuting flows.
We then show that stationary axisymmetric gravity realizes not a single
flow but a \emph{projective} image of the universal kernel, leading to a
unique gravitational trilinear kernel $\mathcal{Y}(\tau_0,\tau_1)$.
Finally, we explain how the Tomimatsu--Sato family and related bilinear
solution sectors arise as degenerate submanifolds of this projected
trilinear structure, consistent with the multilinear analysis
of Ref.~\cite{Fukuyama:2025TS}.

%%%%%%%%%%%%%%%%%%%%%%%%%%%%%%%%%%%%%%%%%%%%%%%%%%%%%%%%%%%%
\section{From bilinear to trilinear: universal hierarchy}

\begin{figure}[t]
\centering
\includegraphics[width=0.65\textwidth]{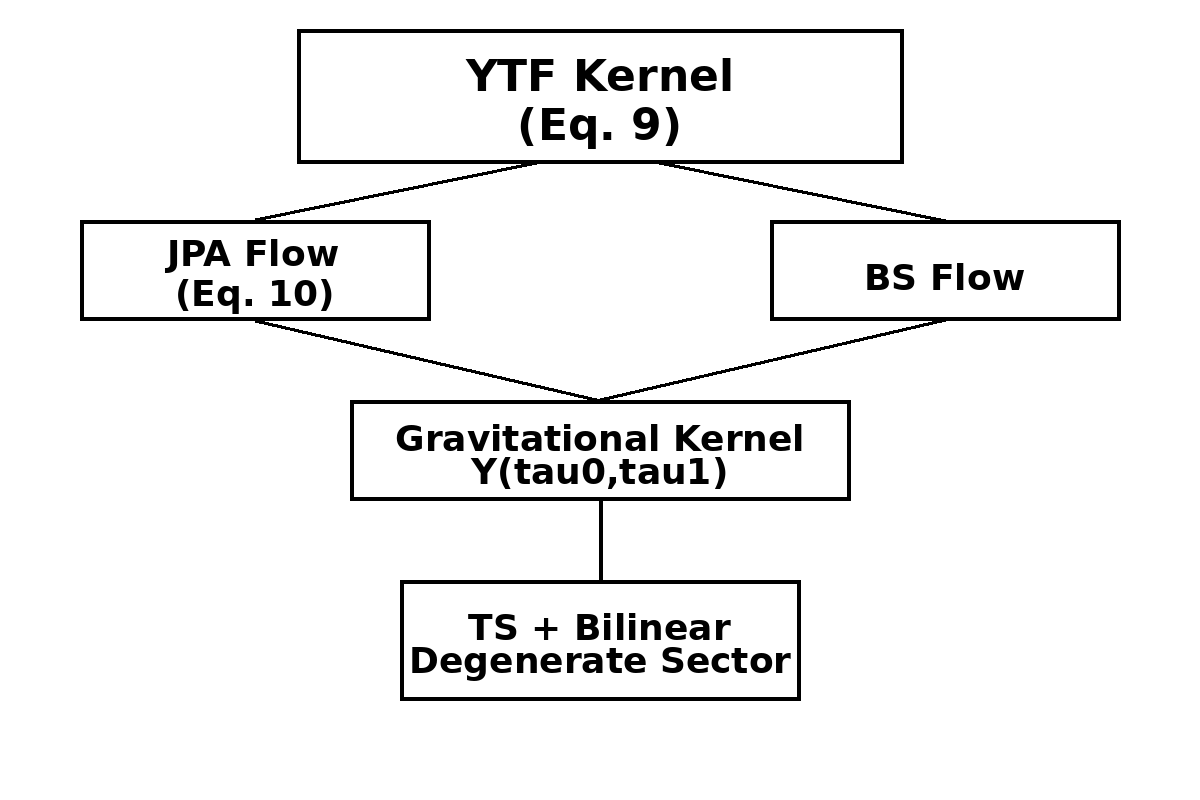}
\caption{
Schematic structure of the YTSF trilinear hierarchy and its gravitational
projection.
The universal trilinear kernel (Eq.~9) generates various multidimensional
integrable equations by selecting particular commuting flows, such as the
JPA flow (Eq.~10), which is the $(3+1)$--dimensional trilinear equation
derived in Ref.~\cite{YTSF1998}, and the
Bogoyavlenskii--Schiff (BS) flow \cite{Bogoyavlenskii1990, Schiff1992}.
Stationary axisymmetric gravity does not correspond to any single flow but
to the projective gravitational kernel $\mathcal{Y}(\tau_0,\tau_1)$ obtained
by projecting the universal kernel onto the $\GL(2)$-covariant Ernst
representation.
The Tomimatsu--Sato family \cite{Tomimatsu1972} and related bilinear solutions
arise as degenerate submanifolds of this projected trilinear structure.
}
\label{fig:scheme-eq55}
\end{figure}

The inadequacy of bilinear structures in higher dimensions is made
explicit by the extension pattern discovered in Ref.~\cite{YTSF1998}.
As summarized in Fig.~\ref{fig:scheme-eq55}, the space of integrable equations
is organized into a hierarchy of increasing structural generality,
ranging from bilinear systems to a universal trilinear kernel and its
projective gravitational realization, in which gravity emerges as a
GL$(2)$-covariant reduction rather than a single commuting flow.

On the bilinear branch, KdV extends to the KP equation,
\begin{equation}
(D_x^4+3D_xD_t-4D_y^2)\,\tau\cdot\tau=0,
\label{KP}
\end{equation}
which is still governed entirely by pairwise interference.
On the other hand, KdV admits a different extension to the
Bogoyavlenskii--Schiff (BS) equation, which already requires trilinear
structures.
The structure summarized in Fig.~1 reflects a hierarchy of increasing
generality.
At the top stands the universal YTF trilinear kernel
(Eq.~9),
which contains all independent multidimensional flows.
Particular trilinear equations, such as the Eq.~10 and the
Bogoyavlenskii--Schiff sector, arise as specific flow selections of this
kernel.
Stationary axisymmetric gravity corresponds not to any single flow but to
a projective realization of the kernel, while the Tomimatsu--Sato family
and related bilinear solutions appear as degenerate reductions.

%%%%%%%%%%%%%%%%%%%%%%%%%%%%%%%%%%%%%%%%%%%%%%%%%%%%%%%%%%%%
\subsection{Trilinear operators and $\mathbb{Z}_3$ projection}

Let $\omega$ be a cubic root of unity, $\omega^3=1$, $1+\omega+\omega^2=0$.
The trilinear Hirota derivative is defined by
\begin{equation}
T_x(a,b,c)=
(\partial_{x_1}+\omega\partial_{x_2}+\omega^2\partial_{x_3})
a(x_1)b(x_2)c(x_3)\Big|_{x_i=x},
\label{Tx}
\end{equation}
with a conjugate $T_x^\ast$ obtained by $\omega\leftrightarrow\omega^2$.
Because $1+\omega+\omega^2=0$, all pairwise contributions are projected
out: $T_x(a,a,b)=0$, so $T_x$ detects only genuine three--way interference.

From now on we specialize to $a=b=c=\tau$ and write
$T_x\tau\cdot\tau\cdot\tau\equiv T_x(\tau,\tau,\tau)$.

%%%%%%%%%%%%%%%%%%%%%%%%%%%%%%%%%%%%%%%%%%%%%%%%%%%%%%%%%%%%
\subsection{The Yu--Toda--Fukuyama trilinear kernel and the generation of flows}

The central object of the Yu--Toda--Fukuyama (YTF) construction is not a particular evolution
equation but a \emph{kernel} that generates an entire integrable hierarchy.
By a kernel we mean a homogeneous trilinear constraint acting on
$\tau\cdot\tau\cdot\tau$ before any specific choice of commuting flows
(``times'') is made.
Different multidimensional integrable equations are obtained from this
kernel by selecting particular flow directions and imposing reductions
among them.

Let $T_x,T_x^\ast,T_y,T_y^\ast,\ldots$ be the trilinear Hirota operators
defined in Sec.~2.1.
A general trilinear kernel is a polynomial $\mathcal{K}$ of these operators
acting on three identical copies of the same $\tau$--function,
\begin{equation}
\mathcal{K}\bigl(T_x,T_x^\ast,T_y,T_y^\ast,\ldots\bigr)\;
\tau\cdot\tau\cdot\tau = 0 .
\label{eq:kernel-def}
\end{equation}
Homogeneity under $\tau\to\lambda\tau$ ensures that
\eqref{eq:kernel-def} defines a hierarchy rather than a single equation.

Yu, Toda and Fukuyama showed that imposing $\mathbb{Z}_3$ symmetry,
homogeneity under $\tau\to\lambda\tau$, and minimal total differential
weight uniquely fixes the trilinear kernel.
The result is Eq.~(56) of Ref.~\cite{YuTodaSasaFukuyama:1998hierarchy}, which
we rewrite here as
\begin{equation}
\Bigl(
36\,T_x^{2}T_t
+T_x^{4}T_z^{\ast}
+8\,T_x^{3}T_x^{\ast}T_z
+9\,T_z^{3}
\Bigr)\,
\tau\cdot\tau\cdot\tau = 0 .
\label{eq:YTSF56}
\end{equation}
Equation (9) is the YTF trilinear kernel.
It generates the Yu-Toda-Sasa-Fukuyama (YTSF) hierarchy of commuting trilinear flows by selecting
particular combinations of the independent variables.

The defining property of the kernel is that it contains several independent
flow directions simultaneously.
In Eq.~\eqref{eq:YTSF56} the $t$--flow appears through the term
$T_x^2T_t$, while the $z$--flow appears through $T_z$, $T_z^\ast$ and the
purely cubic contribution $T_z^3$.
Before any reduction is imposed, all these flows are on an equal footing.
Choosing a particular set of commuting flows (and possibly relating them
by constraints) produces concrete trilinear evolution equations, which are
called \emph{flows} of the hierarchy.

For comparison, the trilinear equation most frequently quoted in the
literature is Eq.~(57) of Ref.~\cite{YTSF1998},
\begin{equation}
\Bigl(
T_x^4 T_z^\ast
+8\,T_x^3 T_x^\ast T_z
-36\,T_x^2 T_t
+27\,T_x T_y^2
\Bigr)\,
\tau\cdot\tau\cdot\tau = 0 .
\label{eq:JPA57}
\end{equation}
Equation (10) is one representative flow of the YTSF hierarchy,
corresponding to the $(3+1)$--dimensional trilinear equation discovered in
Ref.~\cite{YTSF1998}.
Compared with the kernel \eqref{eq:YTSF56}, the trilinear equation
\eqref{eq:JPA57} exhibits a crucial structural restriction.
The kernel \eqref{eq:YTSF56} contains simultaneously several independent
flow generators: a $t$--flow through $T_x^2T_t$, a $z$--flow through
$T_z$, $T_z^\ast$, and in particular a purely cubic dispersion term
$T_z^3$.
These terms encode the full $(3+1)$--dimensional trilinear phase space of
the hierarchy.

By contrast, in \eqref{eq:JPA57} the cubic $z$--sector $T_z^3$ is absent,
and instead a $y$--flow term $T_xT_y^2$ appears.
This is not an independent structure, but corresponds to a reduction of
the kernel in which the $z$--flow is eliminated and replaced by a
constraint linking $z$ to $y$.
In other words, \eqref{eq:JPA57} is obtained from the kernel by imposing
relations among the commuting flows.

This can be seen explicitly as follows.
The kernel \eqref{eq:YTSF56} is a homogeneous constraint of Eq.~\ref{eq:kernel-def}, 
in which $t$ and $z$ are on an equal footing as independent evolution
parameters.
If one now imposes a reduction such as
\[
\partial_z \;\sim\; \partial_y ,
\]
so that the $z$--flow is no longer independent but slaved to a new
variable $y$, the cubic term $T_z^3$ is replaced by a mixed term
proportional to $T_xT_y^2$.
Under such a reduction, the kernel \eqref{eq:YTSF56} reduces precisely to
\eqref{eq:JPA57}.

Thus Eq.~\eqref{eq:JPA57} is not a new kernel but a \emph{particular flow}
of the YTF kernel \eqref{eq:YTSF56}, obtained by freezing one of the
independent directions of the full $(3+1)$--dimensional hierarchy.

The hierarchy therefore closes at the trilinear level for a concrete
reason.
The $\mathbb{Z}_3$--projected trilinear operators already provide a
complete basis of independent cubic dispersions.
Any quadrilinear operator would either violate homogeneity under
$\tau\to\lambda\tau$ or reduce, after projection, to products of these
trilinear blocks.
Hence no new independent flow directions appear beyond the trilinear
kernel \eqref{eq:YTSF56}.

In this precise sense, trilinearity is not merely a minimal extension of
Hirota's bilinear calculus, but the maximal closed algebra of independent
multidimensional soliton flows compatible with the $\mathbb{Z}_3$
projection.

%%%%%%%%%%%%%%%%%%%%%%%%%%%%%%%%%%%%%%%%%%%
\section{Gravitational projection of the trilinear hierarchy}
\label{sec:gravity}
This section corresponds to the third level of Fig.~1.
Here we show how the universal YTF kernel and its flows are projected onto
the $\GL(2)$--covariant gravitational kernel $\mathcal Y(\tau_0,\tau_1)$
appropriate to the Ernst representation.

In Sec.~2.2 the universal trilinear kernel of the YTSF hierarchy was
identified with Eq.~\eqref{eq:YTSF56}. This kernel is written for a single $\tau$--function and treats all flow
directions $(x,t,z,\ldots)$ symmetrically.
Stationary axisymmetric gravity, however, is intrinsically projective:
the physical field is the Ernst potential
$\mathcal{E}=\tau_1/\tau_0$, and the fundamental variables are the
$\GL(2)$--equivalence class of pairs $(\tau_0,\tau_1)$.
Therefore, the YTF kernel cannot be used directly; it must be projected
onto a $\GL(2)$--covariant trilinear form acting on $(\tau_0,\tau_1)$.

\subsection{Projective structure of the Ernst representation}

The projective nature of the Ernst formulation implies the invariance
\[
(\tau_0,\tau_1)\;\longmapsto\;
(a\tau_0+b\tau_1,\;c\tau_0+d\tau_1),\qquad ad-bc\neq 0,
\]
under which $\mathcal{E}=\tau_1/\tau_0$ is unchanged.
Any trilinear expression intended to represent the gravitational dynamics
must therefore be $\GL(2)$ covariant and homogeneous under
$(\tau_0,\tau_1)\to\lambda(\tau_0,\tau_1)$.

The universal kernel \eqref{eq:YTSF56} already satisfies homogeneity in a
single $\tau$, but it does not respect this projective structure.
We must therefore construct from it a projected kernel that acts on
$(\tau_0,\tau_1)$ and is insensitive to $\GL(2)$ redefinitions.

\subsection{Construction of the gravitational kernel}
The trilinear kernel introduced in Sec.~2 is purely algebraic: it encodes
the universal structure of multidimensional soliton hierarchies in terms
of a single $\tau$--function.
Stationary axisymmetric gravity, however, is intrinsically different.
Its fundamental variable is not a single $\tau$ but the projective ratio
$\mathcal{E}=\tau_1/\tau_0$, and its dynamics is governed by the Ernst
equation.

The central question is therefore not how to apply the YTF kernel to a
single $\tau$, but how to project it onto the $\GL(2)$--equivalence class
$(\tau_0,\tau_1)$ in a way that reproduces the nonlinear geometry of
gravity.
This projection is highly nontrivial, because it must simultaneously
respect
(i) projective $\GL(2)$ covariance,
(ii) homogeneity,
and (iii) the intrinsic nonlinearity of the Ernst equation.

In Ref.~\cite{Fukuyama:2025TS}, it was shown that the Ernst equation admits
a multilinear $\tau$--function representation in which its nonlinearity
is encoded not bilinearly but through a trilinear structure.
This observation provides the physical motivation for introducing a
gravitational trilinear kernel:
it is the unique $\GL(2)$--covariant projection of the universal YTF kernel
that reproduces the Ernst dynamics.

The projection is achieved by replacing the single $\tau$ in the YTF
kernel by the projective pair $(\tau_0,\tau_1)$ in the minimal
$\GL(2)$--covariant way.
The unique antisymmetric combination of three $\tau$--copies with minimal
weight is
\begin{equation}
\tau_1\,T(\tau_0,\tau_0,\tau_1)-\tau_0\,T(\tau_1,\tau_1,\tau_0),
\label{eq:proj-structure}
\end{equation}
where $T$ stands for any of the trilinear operators $T_x,T_y,\ldots$.
This structure vanishes for proportional pairs
$(\tau_0,\tau_1)\propto(1,1)$ and therefore measures the nontrivial
projective content of the Ernst field.

Applying this projection to the YTF kernel \eqref{eq:YTSF56} leads to the
gravitational trilinear kernel
\begin{align}
\mathcal{Y}(\tau_0,\tau_1)
&=
\partial_x\!\Bigl[
\tau_1 T_x(\tau_0,\tau_0,\tau_1)
-\tau_0 T_x(\tau_1,\tau_1,\tau_0)
\Bigr]
\nonumber\\
&\quad+
\partial_y\!\Bigl[
\tau_1 T_y(\tau_0,\tau_0,\tau_1)
-\tau_0 T_y(\tau_1,\tau_1,\tau_0)
\Bigr].
\label{eq:Ykernel}
\end{align}
This is the $\GL(2)$--covariant projection of the universal kernel
\eqref{eq:YTSF56} and will be referred to as the \emph{gravitational kernel}.
\subsection{Relation to the Ernst equation}
The role of $\mathcal{Y}(\tau_0,\tau_1)$ is to encode the intrinsic
nonlinearity of the Ernst equation in trilinear form.
When $\mathcal{Y}(\tau_0,\tau_1)=0$ is imposed together with
$\mathcal{E}=\tau_1/\tau_0$, one recovers the stationary axisymmetric
Einstein equations in the Ernst form.

In particular, the Ernst equation does not single out any preferred
evolution variable among the hierarchy of commuting flows.
It is therefore not a flow of the YTSF hierarchy, but a constraint
obtained by projecting the full trilinear kernel onto the
$\GL(2)$--covariant projective variables $(\tau_0,\tau_1)$.

Conceptually, the universal kernel \eqref{eq:YTSF56} describes the full
multidimensional trilinear hierarchy, while the gravitational kernel
\eqref{eq:Ykernel} represents its projective realization appropriate to
stationary gravity.
In this sense gravity does not correspond to a particular flow such as
\eqref{eq:JPA57}, but to the projection of the entire kernel.

\section{Tomimatsu--Sato sector as a degenerate reduction}
\label{sec:TS}
This section corresponds to the fourth level of Fig.~1.
We show that the Tomimatsu--Sato (TS) family \cite{Tomimatsu1972} and related bilinear solutions arise
as degenerate submanifolds of the gravitational kernel
$\mathcal Y(\tau_0,\tau_1)$.

The TS family provides a concrete realization of how the
universal trilinear hierarchy collapses to a bilinear structure when strong
geometric constraints are imposed.
In Ref.~\cite{Fukuyama:2025TS} the $\tau$--function formulation of the TS
spacetimes was analyzed in detail, and it was shown that the entire TS
family is generated by Hirota bilinear equations.
From the present viewpoint, this remarkable fact admits a precise
structural interpretation.

The universal YTSF hierarchy is governed by the trilinear kernel
\eqref{eq:YTSF56}.
Stationary gravity corresponds not to a particular flow of this hierarchy
but to its projective realization through the gravitational kernel
$\mathcal{Y}(\tau_0,\tau_1)$ defined in Sec.~\ref{sec:gravity}.
For generic $(\tau_0,\tau_1)$ this kernel encodes genuine three--way
interference among independent flow directions.
However, in the TS sector the $\tau$--pair is constrained so strongly by
axisymmetry and rational Ernst potentials that this three--way
interference becomes trivial.

More precisely, the TS $\tau$--functions satisfy algebraic relations that
force the projected kernel to vanish identically,
\begin{equation}
\mathcal{Y}(\tau_0,\tau_1)\equiv 0
\qquad \text{(TS sector)}.
\label{eq:TS-degenerate}
\end{equation}
This identity expresses the collapse of the trilinear hierarchy to a
bilinear one.

Indeed, Ref.~\cite{Fukuyama:2025TS} shows that the TS solutions obey a closed
bilinear hierarchy of the form
\begin{equation}
\mathcal{B}_{\rm TS}(D_x,D_y,\ldots)\,\tau\cdot\tau=0,
\label{eq:TS-bilinear}
\end{equation}
which reproduces the full TS family.
Equation \eqref{eq:TS-bilinear} should be viewed not as an independent
structure but as a reduction of the trilinear kernel: the TS sector lies on
a lower--dimensional submanifold of the universal YTSF phase space where the
three--way interference encoded in \eqref{eq:YTSF56} disappears.

Physically, this means that Tomimatsu--Sato spacetimes probe only a highly
restricted region of the full trilinear hierarchy.
They are analogous to special reductions of the KP hierarchy, such as KdV,
which admit a bilinear description even though the full KP structure is
multidimensional.
The TS family therefore provides a nontrivial but degenerate realization of
the universal YTSF hierarchy within stationary gravity.

This interpretation resolves the apparent tension between the success of
bilinear methods in generating TS solutions and the necessity of trilinear
structures in the general gravitational theory.
Bilinear equations work in the TS sector precisely because the
gravitational kernel $\mathcal{Y}(\tau_0,\tau_1)$ vanishes there; beyond
this degenerate locus, genuinely trilinear interactions are unavoidable.
We emphasize that the TS family does not exhaust the class of
solutions that admit a bilinear description.
As discussed in Ref.~\cite{Fukuyama:2025TS}, there exist further stationary
axisymmetric solutions, beyond the standard TS family, which are also
generated by a bilinear $\tau$--function hierarchy.
From the present viewpoint, these solutions correspond to other degenerate
reductions of the universal trilinear kernel.
They lie on larger but still lower--dimensional submanifolds of the YTSF
phase space, where the gravitational kernel $\mathcal{Y}(\tau_0,\tau_1)$
either vanishes or becomes reducible to bilinear form.

This observation suggests a stratified structure of stationary solutions:
the full YTSF hierarchy at the top, its projective gravitational image in
the middle, and a family of bilinear submanifolds—including but not limited
to the Tomimatsu--Sato sector—at the bottom.

%%%%%%%%%%%%%%%%%%%%%%%%%%%%%%%%%%%%%%%%%%%%%%%%%%%%%%%%%%%%
\section{Conclusion}

We have shown that the trilinear kernel of Yu--Toda--Fukuyama,
Eq.~\eqref{eq:YTSF56}, provides the universal master constraint underlying
multidimensional integrable hierarchies.
Unlike the more frequently quoted trilinear equation
Eq.~\eqref{eq:JPA57}, which represents a
specific flow, the kernel \eqref{eq:YTSF56} contains all independent flow
directions on an equal footing and generates particular evolution
equations by suitable reductions.

Stationary axisymmetric gravity does not correspond to a single flow of
this hierarchy but to a projective realization of the full kernel.
By imposing $\GL(2)$ covariance, homogeneity, and minimal weight on the
YTF kernel, we obtained a unique gravitational trilinear kernel
$\mathcal{Y}(\tau_0,\tau_1)$.
The Ernst equation is recovered as the vanishing of this projected kernel,
showing that the intrinsic nonlinearity of gravity is encoded in a
trilinear, rather than bilinear, structure.

The Tomimatsu--Sato family was shown to constitute a degenerate reduction
of this universal framework.
In this sector the gravitational kernel $\mathcal{Y}(\tau_0,\tau_1)$
vanishes identically, and the trilinear hierarchy collapses to a bilinear
one, in agreement with the detailed analysis of
Ref.~\cite{Fukuyama:2025TS}.
This clarifies both the power and the limitation of Hirota bilinear methods
in gravitational applications.

These results establish a coherent structural link between the
Yu--Toda--Fukuyama trilinear kernel, its Yu--Toda--Sasa--Fukuyama hierarchy
of flows, stationary axisymmetric gravity, and the Tomimatsu--Sato
solutions, and highlight the fundamental role of trilinear kernels in
multidimensional integrable systems with projective geometry.

%%%%%%%%%%%%%%%%%%%%%%%%%%%%%%%%%%%%%%%%%%%%%%%%%%%%%%%%%%%%
\noindent
{\bf Acknowledgments}  
%The author thanks K.~Morikawa and A.~Tatekawa for illuminating discussions
%on cosmological non-equilibrium dynamics, and the ALICE Collaboration for
%stimulating this work.
This work is supported in part by 
  Grant-in-Aid for Science Research from the Ministry of Education, Science and Culture No.~25H00653.
%%%%%%%%%%%%%%%%%%%%%%%%%%%%%%%%%%%%%%%%%%%%%%%%%%%%%%%%%%%%

%%%%%%%%%%%%%%%%%%%%%%%%%%%%%%%%%%%%%%%%%%%%%%%%%%%%%%%%%%%%

\end{document}